\documentclass{article}

\usepackage{arxiv}
\usepackage{multirow}
\usepackage{hyperref}
\usepackage[utf8]{inputenc} 
\usepackage[T1]{fontenc}    
\usepackage{hyperref}       
\usepackage{url}            
\usepackage{booktabs}       
\usepackage{amsfonts}       
\usepackage{nicefrac}       
\usepackage{microtype}      
\usepackage{lipsum}
\usepackage{graphicx}
\graphicspath{ {./images/} }

\title{A Self-supervised Approach for Semantic Indexing in the Context of COVID-19 Pandemic}

\author{
 Nima Ebadi \\
  Department of Electrical and Computer Engineering\\
  University of Texas at San Antonio\\
  San Antonio, SA 78249 \\
  \texttt{nima.ebadi@utsa.edu} \\
  \And
 Peyman Najafirad \\
  Department of Information Systems and Security \\
  University of Texas at San Antonio\\
  San Antonio, TX 78249 \\
  \texttt{peyman.najafirad@utsa.edu} \\
}

\begin{document}
\maketitle

\begin{abstract}
The pandemic has accelerated the pace at which COVID-19 scientific papers are published. In addition, the process of manually assigning semantic indexes to these papers by experts is even more time-consuming and overwhelming in the current health crisis. Therefore, there is an urgent need for automatic semantic indexing models which can effectively scale-up to newly introduced concepts and rapidly evolving distributions of the hyperfocused related literature. In this research, we present a novel semantic indexing approach based on the state-of-the-art self-supervised representation learning and transformer encoding exclusively suitable for pandemic crises. We present a case study on a novel dataset that is based on COVID-19 papers published and manually indexed in PubMed. Our study shows that our self-supervised model outperforms the best performing models of BioASQ Task 8a by micro-F1 score of 0.1 and LCA-F score of 0.08 on average. Our model also shows superior performance on detecting the supplementary concepts which is quite important when the focus of the literature has drastically shifted towards specific concepts related to the pandemic. Our study sheds light on the main challenges confronting semantic indexing models during a pandemic, namely new domains and drastic changes of their distributions, and as a superior alternative for such situations, propose a model founded on approaches which have shown auspicious performance in improving generalization and data efficiency in various NLP tasks. We also show the joint indexing of major Medical Subject Headings (MeSH) and supplementary concepts improves the overall performance.
\end{abstract}

\section{INTRODUCTION AND BACKGROUND}
\label{sec:introductionandbackgrounds}
To facilitate literature search and storage, curators at National Library of Medicine (NLM) annotate every article with a set of concepts from established categorical semantic terminologies.\cite{tsatsaronis2015overview} This annotation process of scientific articles is generally referred to as semantic indexing. Nevertheless, the manual process of biomedical semantic indexing is time-consuming and financially expensive.\cite{huang2011recommending, mork2013nlm} Therefore, several automated semantic indexing models have been proposed in the literature, including NLM's official Medical Text Indexing tool (MTI).\cite{jin2020tackling,peng2016deepmesh,jin2018attentionmesh,xun2019meshprobenet,segura2016labda,zavorin2016using}

A pandemic situation, however, is an extreme scenario which highlights the importance of automated semantic indexing as researchers desperately require a well compartmentalized database to gain insights about the recent findings.\cite{shokraneh2020lessons} During the current pandemic so many related papers are being published at a much faster pace,\cite{esteva2020co} and the focus of the literature has drastically shifted towards COVID-19 related topics and subtopics,\cite{roberts2020trec} some of them have not had a standard name until a couple of month ago. \cite{rao2020retweets} Such conditions cause challenges for the automatic semantic indexing systems which are based on substantial supervisions and hand-coded features. Albeit the importance of semantic indexing in the pandemic situation, there is a lack of study on the performance of such automated models on the rapidly evolving corpus of COVID-19 related documents.\cite{shokraneh2020lessons} In this research, we present a case study on the state-of-the-art semantic indexing models in the context of COVID-19 pandemic. We analyze the key challenges of these models performing various evaluation (training and testing) schema. We find out the key aspects of the pandemic causing challenges for automatic semantic indexing models are the abrupt changes in the distribution of these indexes, rapid growth of specific topics regarding few indexes from a relatively large set of indexes, and lack of standard terms for newly introduced topics. 

In this research, we attempt to tackle the problem of semantic indexing exclusively in the pandemic situation. We propose a novel semantic indexing methodology suitable for the aforementioned challenges, i.e. that is able to effectively scale-up to COVID-19 literature. Inspired by the state-of-the-art performance of self-supervised learning (SSL) models in various NLP,\cite{lan2019albert,liu2019roberta} and BioNLP,\cite{lee2020biobert} tasks--specifically their generalization and data efficiency capabilities--as well as the best performing models in BioASQ Task 8a,\cite{peng2016deepmesh,jin2018attentionmesh} we design our methodology based on transformers encoding and attention mechanism between the document and candidate indexes. Our experimental results denote our model as a superior alternative over the best-performing models of BioASQ challenge during health crisis situations, like the current one. The main contributions of this study are as follows:

\begin{enumerate}
    \item  We propose a novel semantic indexing approach which can effectively scale up to new distributions, thereby suitable for emergency situations like the current pandemic where the related literature is rapidly evolving. 
    \item Our study bring attention to the main challenges confronting semantic indexing models in the pandemic crises, and attempts to address them by proposing a novel model inspired by the best-performing models of BioASQ challenge, but unlike them, able to leverage self-supervised representation learning and transformer language model to improve efficiency and generalization.
    \item We present a case study on a novel semantic indexing dataset that is based on the COVID-19 related research articles published and manually indexed in PubMed. We use flat and hierarchical measures to evaluate the performance of our model along with the state-of-the-art benchmarks. Our study demonstrate the superiority of our self-supervised approach in scaling to the novel pandemic situation with the relatively small amount of labeled data available.
    \item We also discuss the importance of more fine-grained categorization of documents to supplementary concepts, and show their indexing can actually improve the MeSH indexing performance when performed simultaneously. In addition to major MeSH indexing, we evaluate the performance of simultaneous indexing of both major MeSH and supplementary concepts. 
    \item This paper aims to offer some aid in the process of semantic indexing of the novel COVID-19 literature so as to lighten the load on NLM indexers.
\end{enumerate}

\subsection{Biomedical Semantic Indexing}
Biomedical literature has been collected by the National Library of Medicine (NLM) for the last 150 years. As of 2020, PubMed database contains about 30 Million biomedical journal citations. This number has risen from 12 Million citations in 2004 to 30 Million citations in 2020 having a growth rate of 4\% per year. Through a laborious process, NLM curators fully examine every document and annotate it with a set of hierarchically-organized terminologies developed by NLM called Medical Subject Headings (MeSH\footnote{\scriptsize{{https://www.nlm.nih.gov/mesh/meshhome.html}}}) along with supplementary concepts for more fine-grained categorization.\cite{papagiannopoulou2016large} In 2019, more than 900K biomedical citations were added to PubMed and manually indexed to more than 29K MeSH concept categories\footnote{\scriptsize{{https://www.nlm.nih.gov/pubs/techbull/mj18/brief/mj18\_updates\_2018\_baseline\_stats.html}}}.

In the light of the size and growth rate of such databases, several automated models have been developed to improve the time-consuming and financially expensive process of biomedical semantic indexing through annual competitions such as BioASQ Task a,\cite{283} and presented models, \cite{jin2018attentionmesh,peng2016deepmesh} as well as other BioNLP research venues.\cite{muller2017livivo,zavorin2016using,xun2019meshprobenet} These approaches are either based on i) simple retrieval systems; such as SNOKES team which participated in BioASQ 6th and uses search engine methods along with UMIA concept extractor,\cite{nentidis2017results} Iria another participating team which combines ensemble of the best performing models from previous years challenges with k-NN MeSH masking algorithms,\cite{ribadascole} Segura et al. utilize ElasticSearch to manage the "scalability" issue of the task and the enhanced NLM Medical Text Indexing (MTI),\cite{segura2016labda} Zavorin et al. combines L2R with Medical Text Indexing;\cite{zavorin2016using} or ii) deep learning models with substantial hand-coded features and supervision. DeepMesh which is the best performing model of a couple of edition of BioASQ challenge that combines document to vector models with crafted features from the document and MeSH indexes along with ensemble models fed by those features. Other deep learning approaches include UIMA concept extractor links,\cite{peng2016deepmesh} and AUTH that also uses document to vector approach with an ensemble of machine learning classifier (SVM) fed with document-MeSH features.\cite{papagiannopoulou2016large} Jin et al. and Xun et al. combined retrieval systems with deep recurrent neural networks and attention mechanism and also provide explainability for MeSH indexing decisions.\cite{jin2018attentionmesh,xun2019meshprobenet} The amount of hand-crafted features and supervision required for these models make it difficult for them to effectively scale-up as the biomedical databases do during pandemic crises.\cite{foroughi2020high}

\subsection{Semantic Indexing in Pandemic}

These semantic indexing models are proposed to perform well in normal situations, when there is no specific interest towards specific concepts, and are evaluated based on their overall performance on all major MeSH indexes. \cite{nentidis2019results} In the pandemic situation, however, the focus of the literature has drastically shifted towards the specific concepts and sub-concepts related to the current Coronavirus disease. The number of published documents related to Coronavirus have risen from to a few articles per month to more than 10K articles in June 2020--roughly 1 out of every 11.5 citations are about Coronavirus these days.\cite{RN12503} The rapidly growing and evolving literature of COVID-19 causes challenges for automatic semantic indexing models.\cite{shokraneh2020lessons} Previously introduced semantic indexing models are based on supervised learning approaches and heavily hand-coded features; therefore, they require significant amount of labeled data regarding a specific concept to show decent performance in indexing related documents, and have challenges to effectively scale-up to newly introduced terminologies and sub-concepts; thereby, not suitable for emergency situations like the ongoing health crisis.

On the other hand, self-supervised learning (SSL), where a model is initially trained on a data-rich unsupervised pretext task then fine-tuned on a downstream task, has recently emerged as an effective technique in almost every deep learning problem ranging from computer vision,\cite{tung2017self,misra2020self} NLP,\cite{yang2019xlnet,liu2019roberta,ravanelli2020multi} to Bioinformatics,\cite{masood2015self,noorbakhsh2020deep} and IoT security.\cite{pour2020data,manginointernet,parra2020detecting} Self-supervised learning is known to enhance data efficiency,\cite{chen2020simple}  and generalization,\cite{chen2020adversarial} because the SSL-based model learns some general auxiliary knowledge from the pre-text task that allows the model to "understand" the downstream task better.\cite{raffel2019exploring} Therefore, SSL-based models are more robust towards changing domains and scaling up to new distributions.\cite{radford2019language}

In order for a deep SSL algorithm to be effective, the pre-text learning process should be susceptible to downstream learning.\cite{raffel2019exploring} However, as for the proposed semantic indexing models in the literature, either their architecture or the representation learning objective of the downstream task does not allow pre-training in an unsupervised manner that is useful for the downstream learning. 
In DeepMeSH only the TF-IDF vectorization can be updated with unlabeled data, without any word-level representation learning.\cite{peng2016deepmesh} In AttenMeSH word-level encoding of the document can be updated by pre-training the Bi-GRU on an masked language modeling (MLM) pretext task, but these encodings wouldn't be appropriate enough for the downstream task because i) it cannot involve the document-index attention in the encoding, and ii) more sophisticated architectures have been proven to be more effective in this regard such as transformers.\cite{vaswani2017attention}

\section{MATERIALS AND METHODS}
\label{sec:materials}
\begin{figure}
\centering
\small
\includegraphics[scale=0.47]{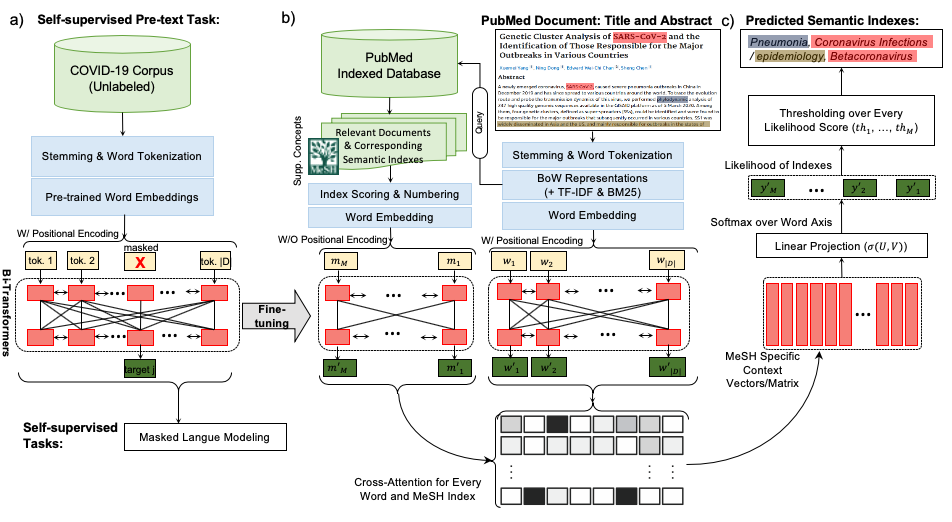}
\caption{\small{Our proposed semantic indexing methodology consists of a self-supervised learning stage (Fig. a) and a fine-tuning stage (Fig. b and c). Fig. a) shows a bi-directional transformers model trained on unlabeled data (i.e. CORD-19) via masked language modeling and next sentence prediction tasks. Fig. b) Every PubMed document along with a set of candidate indexes are encoded via bidirectional transformers self-attention, and cross-attention between every input token/word and index are computed. Fig. c) index specific context vectors (computed by a weighted sum) is passed through a linear projection layer and thresholding process to detect the final semantic indexes which should be assigned to the document. (Example absract is from Yang et al., (2020).\cite{yang2020genetic})}} 
\label{fig:methodology}
\end{figure}

\subsection{Self-supervised Learning Pre-text Task and Document Representations}
\label{sec::self-supervised}
As for the initial representations of documents, we use the word representations provided by BioASQ organizers which is a pre-trained word embedding trained on large-scale corpus of biomedical documents\footnote{\url{http://participants-area.bioasq.org/tools/BioASQword2vec}}. We also use BioASQ provided word tokenizer to parse documents’s title and abstract to a list of the constituting words\footnote{\url{http://participants-area.bioasq.org/tools/}}. We perform stemming and eliminate the stop words as different variants of an individual word or a stop words does not affect the semantic index of a document. 

Afterwards Bag-of-Word representation of each document is computed according to the following equation:
\begin{equation}
\label{equ:bow-input}
    D = [ w_1, w_2, ... w_{|D|}] \hspace{6mm} where \hspace{3mm} w_j \in {\mathbb{R}}^{d_{e_1}}
\end{equation}
Where $|D|$ is the number of non-stopwords in the document, and $d_{e_1}$ is the word embedding size.

For the model to get acquainted with COVID-19 context, we leverage an unlabeled corpus of COVID-19 related documents, called CORD-19,\cite{Wang2020CORD19TC} which is significantly larger than our labeled dataset of indexed documents. We pre-train a bi-directional transformer on masked language modeling (MLM) task with word-tokenized representation of the documents shown in Eq. \ref{equ:bow-input}, along with those of candidate similar indexes\footnote{Note: in the pre-text task we also feed the potentially related indexes for every document to learn latent representations based on the index specific encodings and reinforce the consistency between the pretext and downstream tasks}.\cite{vaswani2017attention,song2019mass}  Such algorithms have shown promising results in many NLP problems regarding deep semantic analysis of scientific documents, such as SciBERT and BioBERT.\cite{beltagy2019scibert,lee2020biobert} In this regard, the document representation $D$ is masked and fed to the transformers model and passed through a positional encoding approach following the original implementation of transformers.\cite{vaswani2017attention} Similar semantic indexes are also fed to the transformer and through a joint document-index attention, index specific encoding of the document is generated. The retrieval and embedding of candidate indexes is discussed in the next section. Next, the transformer model gets trained to predict masked tokens from the index specific encoding (similar to \ref{fig:methodology}.b) but the softmax is applied over the index axis). Unlike BERT and BioBERT, we do not leverage the next sentence prediction (NSP) task for two reasons: 1) sentences ordering is required for QA type of inferences, not semantic indexing and text classifications. 2) It has been proven to be ineffective.\cite{liu2019roberta,lan2019albert}

\subsection{Candidate MeSH Retrieval and Index Representations }
\label{sec::candidate-indexes}
Inspired by Jin et al., as for major MeSH indexes, we initially use a retrieval system to retrieve a subset of related MeSH categories from relevant documents\footnote{We can regard this module as a weak classifier which filter out the negative data which is far more than the positive ones (29K total number of MeSH terms vs. 12.6 terms for every document on average). Jin et al. shows doing so enhances the efficiency and performance of the indexing models as the classifier only focuses on the detection of correct MeSH indexes from a subset of plausible ones.}.\cite{jin2018attentionmesh} Note: we only perform this retrieval process for major MeSH indexes, not for supplementary concepts since there is only 19 of them in COVID-19 dataset.

In this regard, we translate the target biomedical document into a query to extract the relevant ones from the annotated database. We follow the same pre-process of parsing, stemming and stopwords removal of Section \ref{sec::self-supervised}. Every document is represented by their both TF-IDF and BM25 weighted sum of their words, following weighting schems of Wang et al.,\cite{wang2012text} and Paik et al.,\cite{paik2013novel} respectively.

Each document as query is represented as follows:

\begin{equation}
    d = \frac{\sum_{i=1}^{n}{\mathrm{TF-IDF/BM25}}({w}_i,d) \times v_{w_i}}{\sum_{i=1}^{n}{\mathrm{TF-IDF/BM25}}({w}_i,d)}
\end{equation}

Where, $w_i$ is the $i^{th}$ word in document $d$, and $v_{w_i}$ is the word vector from the provided pre-trained embeddings.

Next using cosine similarity scores between the target document and other ones, we find the $K$ relevant documents. Next, we use scoring scheme to re-rank and collect the candidate MeSH indexes. We score every MeSH term by summing their IDF weights in the documents and rank them. The top $M$ with the highest scores are considered for indexing and passed to the next stage.

Semantic indexes representations are quite straight-forward to extract as they are single words. The indexes’ embeddings are as follows:
\begin{equation}
\label{eq:mesh-embedding}
    M = [ m_1, m_2, ... m_{|M|}] \hspace{6mm}  where \hspace{3mm} m_j \in {\mathbb{R}}^{d_{e_2}}
\end{equation}
Where $|M|$ is the number of filtered indexes, and $d_{e_2}$ is the embedding size of every index $m_j$. To simplify the model, we make $d_{e_1} = d_{e_2} = d_{model}$.



\subsection{Index Specific Context Vectors}

After candidate indexes are retrieved, the document BoW representation $D$ along with those of indexes $M$ are fed to the bi-directional transformers which is pre-trained on the self-supervised pre-text task. Positional encoding is performed for documents, not for indexes, to bring in their words' ordering. Initially, $D$ and $M$ are separately encoded to ${D^{'}}$ and ${M^{'}}$ with self-attention mechanism allowing words to only attend other words, and indexes to attend other indexes (there is no cross attention between words and indexes). Self-attention mechanism for ${D^{'}}$ is to capture context-aware representation of words, and for ${M^{'}}$ is to capture correlation and dependencies between indexes which has been shown important. \cite{peng2016deepmesh,sahba2018automatic}

Next, cross-attention between encodings of words and indexes are computed using scaled dot-product attention function, \cite{vaswani2017attention,bendre2020human} as follows:

\begin{equation}
    O = \mathrm{Softmax}(\frac{{M^{'}}{D^{'}}^T}{\sqrt{d_{model}}}){D^{'}} \in {\mathbb{R}}^{|M| \times d_{model}}
\end{equation}

where ${M^{'}}{D^{'}}^T \in {\mathbb{R}}^{|M| \times |D|}$ is the dot product between every index and every word packed together into a matrix multiplication. Softmax is performed over word axis to get attention weights for every index. Finally, $O$ is the index specific context vectors each of which is based on a weighted sum of the word vectors.

\subsection{Projection Layer and Final Prediction}

To compute the likelihood scores of indexes, we apply a linear projection layer with a non-linear activation function $\sigma$ on the context specific vectors $O$ and index encodings $M^{'}$, as following equation:

\begin{equation}
    \hat{Y} = \sigma ({U \cdot {O^T}} + {V \cdot {{M^{'}}^T}} + B)
\end{equation}

where $\hat{Y} \in {\mathbb{R}}^{|M| \times 1}$ is the set of likelihood scores for candidate indexes, and $U,V \in {\mathbb{R}}^{1 \times d_{model}}$ and $B \in \mathbb{R}^{|M| \times 1}$ are trainable parameters. 

Finally, the predicted indexes are computed through thresholding over every likelihood score. Thresholds are defined by maximizing the micro f-measure in the training set, following. \cite{pillai2013threshold}

\section{RESULTS}
\label{sec:results}

\subsection{Dataset}
For self-supervised representation learning (pre-training) stage of our methodology, we use CORD-19 dataset which includes 141K research articles about Coronavirus published in peer-reviewed venues and archival services such as bioRxiv\footnote{\url{https://www.biorxiv.org}} and medRxiv\footnote{\url{https://www.medrxiv.org}}.\cite{Wang2020CORD19TC} These articles are crawled from various medical databases including PubMed's PMC (using the query: COVID-19 and coronavirus research), a COVID-19 corpus maintained by WHO, Elsevier and the aforementioned archival services. A great portion of the dataset (i.e. 48K) of these articles have been published in 2020, in the context of pandemic.

\begin{table}[h!]
\normalsize
\centering
\begin{tabular}{l|c c c}
\hline
\multirow{2}{40pt}{\textbf{Dataset}} & \textbf{No. of} & \textbf{No. of} & \textbf{No. of Supp.} \\ 
 & \textbf{Documents} & \textbf{MeSH Indexes} & \textbf{Concepts} \\ \hline 
\multirow{2}{40pt}{\textbf{COVID SSL}} & \multirow{2}{15pt}{141,764} & \multirow{2}{15pt}{-} & \multirow{2}{15pt}{-} \\ 
& & & \\ \hline
\multirow{2}{40pt}{\textbf{BioASQ}} & \multirow{2}{15pt}{2,501,982} & 27,114 & - \\
     &   & (12.84/doc.) &  \\ \hline
\multirow{2}{40pt}{\textbf{COVID Train}} & \multirow{2}{15pt}{10,210} & 14,335 & 18 \\
     &   & (16.6/doc.) & (1.97/doc.) \\ \hline
\multirow{2}{40pt}{\textbf{COVID Test}} & \multirow{2}{15pt}{3,463} & 10,922 & 15 \\
     &   & (17.9/doc.) & (1.92/doc.) \\ \hline
\end{tabular}
\caption{Data Descriptive for self-supervised representation learning and supervised semantic indexing (training and evaluation) datasets.} 
\label{table:data_desc}
\end{table} 
\begin{figure}
    \centering
    \includegraphics[scale=0.36]{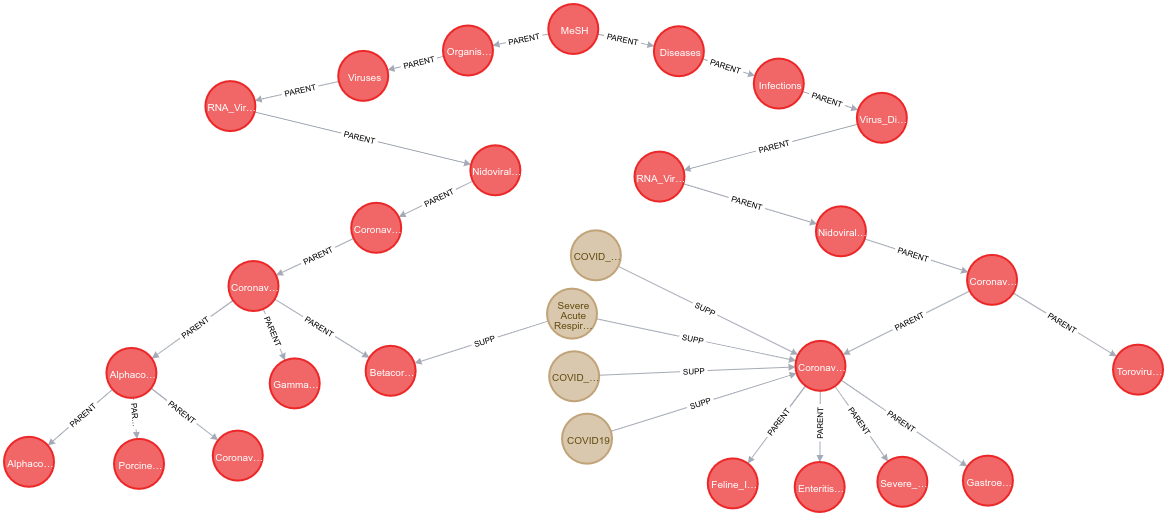}
    \caption{{Semantic Indexes of Coronavirus. The major MeSH and supplementary concepts form a Directed Acyclic Graph (DAG). In this figure, supplementary concepts have been added only for two of the nodes. } }
    \label{fig:mesh-supp-dag}
\end{figure} 
For supervised training we use two sets: i) BioASQ Task 8a Test Sets (from 2015-2019): The training set consists of 2,501,982 annotated articles from PubMed. MeSH labels are manually assigned to the articles by National Library of Medicine indexers. The dataset includes journal name in which the article has been published, article's title and abstract along with MeSH indexes for the training sets. On average, each article is indexed with 12.84 MeSH categories.\cite{283} ii) Recently collected COVID-19 related documents from PubMed: we use 13K latest documents, published and annotated in 2020, related to COVID-19 crawled from PubMed using this query: \textit{covid-19} AND \textit{severe acute repository syndrome 2} AND \textit{sars-cov-2},\cite{RN12503} and evaluation results are calculated by this dataset. Table \ref{table:data_desc} provide information about the statistics of the datasets. We utilize the MeSH majors from both sets as well as the supplementary concept from our set to measure the performance in detail. As shown in Figure \ref{fig:mesh-supp-dag} major MeSH indexes and supplementary concepts form a Directed Acyclic Graph (DAG) where there is a hierarchical relation between two major MeSH (parent-child) and a mapping relation between mesh and supplementary concepts.

\subsection{Experimental Setup}
As for the MeSH retrieval part of our methodology, we use bm25/tf-idf bag-of-words representation methods with vocabulary size of 90K. The retrieval components, i.e. vectorization global features as well as the thresholding features, are trained using the train set only to avoid data bleeding.\cite{riedel2017simple} The bi-directional transformer is implemented using TensorFlow (2.0) Eager,\cite{agrawal2019tensorflow,abadi2016tensorflow} and \textit{tensor-2-tensor}\footnote{\url{https://github.com/tensorflow/tensor2tensor}} library. The hyperparameter values are shown in Table \ref{table:hyperparam}. We use Adam optimizer and early stopping strategies.\cite{yao2007early}  We apply three versions of our methodology: i) base bi-transformer without self-supervised training: where the model is not pre-trained on CORD-19 dataset, section b) and c) of Figure \ref{fig:methodology} with random initialization of the parameters; ii) base bi-transformer with masked language modeling (MLM) as the self-supervised pre-text task; and iii) large bi-transformer with MLM self-supervised tasks. 

\begin{table}[h!]
\normalsize
\centering
    \begin{tabular}{l c}
    \hline
\textbf{Hyperparameter}   &    \textbf{Value(s)}    \\ \hline
$ |V| $        &    900K, 1800K    \\
Number of Retrieved Documents       &  0.1K , 0.5K, \textbf{5K}    \\
Number of MeSH Candidates       &  128, 256, 512, \textbf{1024}    \\
Word Embedding Size for Documents    &   128, \textbf{256}, 512, {\textbf{1024}$^*$} \\ 
Hidden Size       &  128, \textbf{256}, 512, {\textbf{1024}$^*$}  \\ 
Feed Forward       &    \textbf{256}, 384, {\textbf{512}$^*$} \\
Number of Layers       &   \textbf{4}, 5, {\textbf{6}$^*$}, 7, 8    \\ 
Learning Rate   &  0.002, 0.001, \textbf{0.0005}, \textbf{0.0001} \\\hline

    \end{tabular}
    \caption{ Hyperparameters values. We use bold text for the optimal ones among all tried values. $^*$ refer to those for Bi-Trans Large.} 
    \label{table:hyperparam}
\end{table} 

\subsection{Evaluation Metrics}
Following BioASQ challenge, we evaluate the performance of the semantic indexing models based on two sets of evaluation measures: i) flat: Accuracy,Micro and Macro F-measures, and ii) hierarchical: lowest common Ancestor F-measure (LCA-F).

Accuracy is the fraction of correct predictions. However, in multi-label classification problems true and predicted classes could be a set of labels for every example; therefore, there is an additional notion on partially correct. To capture this, precision and recall measures are computed for every class separately. Then, the results are aggregated using micro-averaging and macro-averaging strategies to compute micro (MiP/MiR) and macro precision/recall (MaP/MaR) respectively. Micro-averaging is evaluating the average difference between the predicted labels and the actual labels globally for each test example, and then averaging over all examples in the test set. The second strategy is macro-averaging evaluation in which each label is evaluated separately then averaged over all the labels. Finally the micro and macro f-measures (MiF and MaF) are computed base on the harmonic mean of the corresponding precision and recall. MiF is more affected by the performance of frequent indexes, while MaF treats every index equally.\cite{ebadi2019implicit} Following BioASQ, MiF is the major flat measure in our presented case study.

As shown in Figure \ref{fig:mesh-supp-dag}, semantic indexes have a hierarchical relation between one another. Therefore, in addition to flat measures, hierarchical measures are also used to evaluate the hierarchical classification performance of the semantic indexing models. In this regard, we leverage Lowest Common Ancestor F-measure (LCA-F)  the algorithm provided by Kosmopoulos et al.,\cite{kosmopoulos2015evaluation} which is the same algorithm used in BioASQ challenge\footnote{\url{https://github.com/BioASQ/Evaluation-Measures/tree/master/hierarchical}}. In LCA-F measure, sets of true class and predicted class are compared based on union of their corresponding augmented graphs which encompass all the lowest common ancestors between every pair. The algorithm has shown desirable results in various hierarchical text classification tasks.

\subsection{Major MeSH Indexing}
Table \ref{table:evaluation-major-mesh} shows the performance of the semantic indexing models with the aim of indexing only major subject headings (i.e. major MeSH). The models are trained on BioASQ data from 2015-2019 (excluding the recent COVID-19 documents) as well as COVID training set. They are tested on COVID testing set. 
\begin{table}[h!]
\centering
\scriptsize
\begin{tabular}{l c c c c}
\hline
\textbf{Model}                & \textbf{LCA-F}  & \textbf{Micro F1} & \textbf{Macro-F}  & \textbf{Accuracy} \\ \hline \hline
\textbf{Medical Text Indexer (MTI)} (Default)          & 0.5083 & \textbf{0.6521} & 0.4553 & 0.4267 \\
\textbf{MTI} (First Line Index)        &  0.5011 & 0.6152 & \textbf{0.5038} & 0.4419 \\ 
\textbf{Deep MeSH}                     & \textbf{0.5732} & \textbf{0.6974} & \textbf{0.5024} & 0.4612  \\ 
\textbf{Attention MeSH}                & \textbf{0.5417} & \textbf{0.6571} & 0.4928 & \textbf{0.5147} \\ 
\textbf{iria}                          & 0.4542 & 0.4908 & 0.3491 & 0.2179  \\ 
\textbf{xgx}                           & 0.4266 & \textbf{0.6368} & 0.4934 & \textbf{0.5018 }\\ 
\textbf{MeSHmallow}                    &  0.3751 & 0.5172 & 0.3570 & 0.3916 \\ \hline
\textbf{BioTrans} (Base) (w/o SSL)                      &  0.4899 & 0.6013 & 0.3725 & 0.3309 \\ 
\textbf{BioTrans} (Base) (w/ MLM SSL)                    &  0.5211 & \textbf{0.6588} & 0.4733 & \textbf{0.5096} \\ 
\textbf{BioTrans} (Large) (w/ MLM SSL)                      &  \textbf{0.5683} & \textbf{0.7071} & \textbf{0.5044} & \textbf{0.5319} \\ \hline
\end{tabular}
\vspace{1mm}
\caption{MeSH indexing performances of our deep transformer and self-supervised learning based models along with the state-of-the-art ones in terms of LCA F-measure (hierarchical measure) as well as accuracy, micro and macro F-measure (flat measure).} 
\label{table:evaluation-major-mesh}
\end{table} 
As shown in \autoref{table:evaluation-major-mesh}, semantic indexing models which are based on deep representation learning algorithm demonstrate better performance in scaling up to novel pandemic situation. Our base and large model with self-supervised training perform state-of-the-art beating most of the baselines by 0.1 micro f-measure. The self-supervised training has shown significant improvement over the vanilla version. Large transformer model also shows better capacity to scale up to new concepts.

The higher performance of self-supervised models reveals that the models learn some sort of common sense (acquire general knowledge) about the pandemic and new distribution of major MeSH indexes. 


\subsection{Efficiency w.r.t COVID-19 Training Data}

\begin{figure}[h!]
    \centering
    \scriptsize
    \includegraphics[scale=0.56]{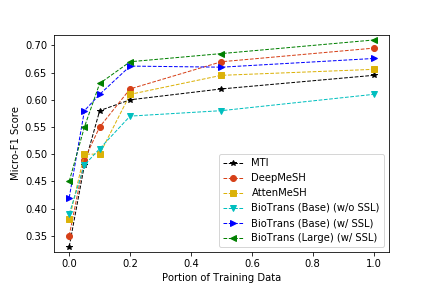}
    \caption{{MeSH indexing performance w.r.t the size of COVID-19 training data based on their Micro-F score. COVID-19 related data is chronologically ordered and then divided; therefore, the horizontal axis is directly related to the date these papers are published.} }
    \label{fig:dataefficiency}
\end{figure}

To evaluate how efficiently the semantic indexing models scale-up to the novel Coronavirus related literature, we chronologically sort the COVID-19 training dataset, and train each model with the following proportions of the data to evaluate their zero- and few-shot performance along with their data efficiency: 0.0 (zero-shot evaluation), 0.05, 0.1, 0.2, 0.5 and 1 (the whole data). 
Figure \ref{fig:dataefficiency} shows the MeSH indexing performance of the top performing models from Table \ref{table:evaluation-major-mesh} based on the size of the exclusive COVID-19 training data. The beginning performance is their zero-shot performance, when the models have only been trained on BioASQ dataset and have not seen a COVID-19 paper yet. The SSL-based versions of our BioTrans model achieve substantially superior performance until almost half of the training data is fed, especially in the very beginning. They reach 0.95 of their optimum performance by only 0.2 of the data. Other models reach this point once half of the training data is fed which means a couple of months delay, an essential issue for a pandemic crisis. 
Our BioTrans model which is not pre-trained on COVID-19 SSL dataset does not learn effectively with the COVID-19 supervised data, and simply follows similar learning speed to those of AttenMeSH and DeepMeSH as it has been inspired by these techniques. This shows that the major strength of using such architecture--bi-directional transformers encoding with attention between documents and indexes--emerges when it undergoes a self-supervised learning process.

\subsection{Indexing of Major MeSH and Supplementary Concepts}

As the literature gets hyper-focused towards specific topics in the context of pandemic, classification to more fine-grained indexes becomes critical. Therefore, we also present evaluation of simultaneous indexing of major Mesh and supplementary concepts. In this regard, the trained models on BioASQ are simply fine-tuned to detect supplementary concepts of COVID training set in addition to the major mesh indexes. Supplementary concepts are added as new classes to the potential indexes.
\begin{table}[h!]
\centering
\scriptsize
\begin{tabular}{l c c c}
\hline
& \multicolumn{3}{c}{\textbf{Micro-F}} \\
\textbf{Model}                & \textbf{Major MeSH}  & \textbf{Supp. Concepts}  & \textbf{Both} \\ \hline 
\textbf{MTI} (Default)          & 0.6188  & 0.7478  & 0.6581 \\
\textbf{MTI} (First Line Index)        &  0.6591 &  0.7247 & 0.6713 \\ 
\textbf{DeepMeSH}                     & \textbf{0.7732} & \textbf{0.8143} & \textbf{0.79011} \\ 
\textbf{AttentionMeSH}                & \textbf{0.7417} & \textbf{0.7991} & 0.7551 \\ 
\textbf{iria}                          & 0.5663 & 0.6127 & 0.5985 \\ 
\textbf{xgx}                           & 0.6202 & 0.7728 & 0.6730 \\ 
\textbf{MeSHmallow}                    &  0.5779 &  0.6187 & 0.6004 \\ \hline
\textbf{BioTrans} (Base) (w/o SSL)                      &  0.6977 &  0.7513 & 0.7199 \\ 
\textbf{BioTrans} (Base) (w/ MLM SSL)                    &  \textbf{0.7283} &  \textbf{0.8092} & 0.7560 \\ 
\textbf{BioTrans} (Large) (w/ MLM SSL)                      &  \textbf{0.7840} &  \textbf{0.8372} & \textbf{0.8067} \\ \hline
\end{tabular}
\vspace{1mm}
\caption{Performance of simultaneous indexing major MeSH and supplementary concepts in the context of COVID-19. Micro F-measure has been calculated for supplementary concepts and major MeSH separately and jointly.} 
\label{table:evaluation-supp-concept}
\end{table}

As demonstrated in Table \ref{table:evaluation-supp-concept} the performance of the baselines is improved by fine-tuning them to detect supplementary concepts as well. This shows the importance of more fine-grained indexing. In comparison to baselines our model improved even more with the aid of supplementary concepts.

\section{CONCLUSION AND DISCUSSION}
\label{sec:conclustion}
In this research, we propose a novel semantic indexing approach based on self-supervised deep representation learning models to tackle this task in the current health crisis. We present a case study on COVID-19 literature collected from recently indexed documents in PubMed. We compare the performance of our model with the state-of-the-art baselines based on flat and hierarchical measures. Our study shows the presented self-supervised model outperforms the baselines with the small amount of labeled data available. We further evaluate the indexing of supplementary concepts along with the major MeSH indexes demonstrating the state-of-the-art performance. We also show that the indexing of supplementary concepts improves MeSH indexing performance of our model explaining the importance of more fine-grained categorization of documents in the current pandemic situation.
In future, we will attempt to continue our case study as COVID-19 documents are being published and indexed in PubMed. We will mainly focus on improving the data efficiency and generalization aspects of semantic indexing models as COVID-19 literature is rapidly evolving. We will also try sophisticated few- and zero-shot learning techniques to better handle newly introduced concepts.



\section*{FUNDING}
The authors gratefully acknowledge the use of the services of Jetstream cloud, funded by National Science Foundation (NSF) awards 1445604, and the Cloud Technology Endowed Professorship.

\bibliographystyle{unsrt}
\bibliography{document}

\end{document}